\documentclass{article}
\usepackage{amssymb}
\usepackage{amsmath}

\input{tcilatex}
\begin{document}

\begin{center}

{\LARGE Diophantine properties of the \textit{zeros} of (monic) polynomials
the \textit{coefficients} of which are the \textit{zeros} of Hermite
polynomials}

\bigskip

$^{\ast }$\textbf{Oksana Bihun}$^{1}$ and $^{+\lozenge }$\textbf{Francesco
Calogero}$^{2}\bigskip $

$^{\ast }$Department of Mathematics, University of Colorado, Colorado
Springs, CO, USA

$^{+}$Physics Department, University of Rome \textquotedblleft La Sapienza''

$^{\lozenge }$Istituto Nazionale di Fisica Nucleare, Sezione di Roma

$^{1}$obihun@uccs.edu

$^{2}$francesco.calogero@roma1.infn.it, francesco.calogero@uniroma1.it

\bigskip

\textit{Abstract}
\end{center}

Let $c_{m},$ with $m=1,...,N$ (with $N$ an arbitrary positive integer, $%
N\geq 2$) be the $N$ zeros (arbitrarily ordered) of the Hermite polynomial $%
H_{N}\left( c\right) ,$ of order $N$ and argument $c$: $H_{N}\left(
c_{m}\right) =0$. Let the monic polynomial $p_{N}\left( z\right) $ of degree 
$N$ in the variable $z$ be defined as follows:%
\begin{equation*}
p_{N}\left( z\right) =z^{N}+\sum_{m=1}^{N}\left( c_{m}~z^{N-m}\right)
=\prod\limits_{n=1}^{N}\left( z-z_{n}\right) ~.
\end{equation*}%
The first equality identifies the $N$ \textit{coefficients} $c_{m}$ of this
polynomial $p_{N}\left( z\right) $ as the $N$ \textit{zeros} $c_{m}$ of the
Hermite polynomial of order $N$; note that there are $N!$ such \textit{%
different} polynomials $p_{N}\left( z\right) $, depending on the ordering
assignment of the $N$ zeros $c_{m}$ of the Hermite polynomial of order $N$.
The second equality identifies (uniquely up to permutations) the $N$ zeros $%
z_{n}$ of the polynomial $p_{N}\left( z\right) $. In this paper we define in
terms of these $N$ zeros $z_{n}$ two $N\times N$ matrices $\mathbf{M}^{(1)}$
and $\mathbf{M}^{(2)}$ with \textit{integer} respectively \textit{%
square-integer} eigenvalues $\lambda _{m}^{(1)}=m$ respectively $\lambda
_{m}^{(2)}=m^{2}$, $m=1,...,N$. The technique whereby these findings are
demonstrated can be extended to other named polynomials.



\section{Introduction}

\textbf{Notation 1.1}. Unless otherwise indicated, hereafter $N$ is an 
\textit{arbitrary positive integer}, $N\geq 2$, indices such as $n,$ $m,$ $%
\ell ,$ $j,$ $k,...$ run over the \textit{integers} from $1$ to $N,$ \textbf{%
boldface} lower-case letters indicate $N$-vectors (for instance the vector $%
\mathbf{z}$ has the $N$ components $z_{n}$) and \textbf{boldface} upper-case
letters indicate $N\times N$ matrices (for instance the matrix $\mathbf{M}$
has the $N^{2}$ components $M_{nm}$). The \textit{imaginary unit} is
hereafter denoted as $\mathbf{i}$ ($\mathbf{i}^{2}=-1;$ of course $\mathbf{i}
$ is not a $N$-vector!). For quantities depending on the independent
variable $t$ a superimposed dot indicates differentiation with respect to $t$%
. The Kronecker symbol $\delta _{nm}$ has the usual meaning: $\delta _{nm}=1$
if $n=m$, $\delta _{nm}=0$ if $n\neq m$. And we adopt throughout the usual
convention according to which a void sum vanishes and a void product equals
unity: $\sum_{j=J}^{K}f_{j}=0,$ $\tprod\nolimits_{j=J}^{K}f_{j}=1$ if $J>K$.
Finally we introduce the following two convenient notations: 
\begin{subequations}
\label{sigma}
\begin{equation}
\sigma _{j}\left( \mathbf{z}\right) =\sum_{1\leq s_{1}<s_{2}<...<s_{j}\leq
N}z_{s_{1}}z_{s_{2}}\cdots z_{s_{j}}~,  \label{sigmam}
\end{equation}%
\begin{equation}
\sigma _{m,j}\left( \mathbf{z}\right) =\sum_{{\footnotesize 1\leq
s_{1}<s_{2}<\ldots <s_{j-1}\leq N~;~s_{k}\neq m,~k=1,...,j-1}%
}z_{s_{1}}z_{s_{2}}\cdots z_{s_{j-1}}~,  \label{sigmamj}
\end{equation}%
where of course the symbol 
\end{subequations}
\begin{equation*}
\sum_{1\leq s_{1}<s_{2}<...<s_{j}\leq N}
\end{equation*}%
denotes the sum from $1$ to $N$ over the $j$ integer indices $%
s_{1},s_{2},\ldots ,s_{j}$ with the restriction that $s_{1}<s_{2}<\ldots
<s_{j}$, while the symbol%
\begin{equation*}
\sum_{{\footnotesize 1\leq s_{1}<s_{2}<\ldots <s_{j-1}\leq N~;~s_{k}\neq
m,~k=1,...,j-1}}
\end{equation*}%
denotes the sum from $1$ to $N$ over the $(j-1)$ indices $s_{1},s_{2},\ldots
,s_{j-1}$ with the restriction that $s_{1}<s_{2}<\ldots <s_{j-1}$ and
moreover the requirement that \textit{all these indices be different from }$m
$. We note that $\sigma _{m,1}(\mathbf{z})=0$ according to the convention
(see above) that a sum over an empty set of indices equals zero. $%
\blacksquare $

A class of properties satisfied by the $N$ zeros $z_{n}$ of several named
polynomials of order $N$ has been obtained via the identification of $%
N\times N$ matrices, constructed with these $N$ zeros $z_{n}$, which feature
eigenvalues displaying remarkable \textit{Diophantine} properties: see for
instance \cite{ABCOP1979, BC}. A general technique to arrive at such results
goes through the following two steps.

\textit{First step}. A dynamical system is manufactured, characterized, say,
by $N$ Newtonian equations of motion (written as follows in $N$-vector
notation), 
\begin{subequations}
\begin{equation}
\mathbf{\dot{\zeta}}=\mathbf{i}~\mathbf{f}\left( \mathbf{\zeta }\right) ~,~~~%
\mathbf{\zeta }=\mathbf{\zeta }\left( t\right) ~,  \label{DS}
\end{equation}%
featuring the following two peculiar properties. (i) This dynamical system
is \textit{isochronous }with period $2\pi $: \textit{all} its solutions are 
\textit{completely periodic} with period $2\pi $,%
\begin{equation}
\mathbf{\zeta }\left( t+2~\pi \right) =\mathbf{\zeta }\left( t\right) ~.
\label{T}
\end{equation}%
(ii) This dynamical system features the equilibrium configuration 
\end{subequations}
\begin{equation}
\mathbf{\zeta }\left( t\right) =\mathbf{z~,}  \label{Equi}
\end{equation}%
where the $N$ components $z_{n}$ of the $N$-vector $\mathbf{z}$ are just the 
$N$ zeros of the named polynomial under consideration.

\textit{Second step}. The behavior of dynamical system (\ref{DS}) in the 
\textit{infinitesimal} neighborhood of its equilibrium (\ref{Equi}) is
investigated in the standard manner, by setting%
\begin{equation}
\mathbf{\zeta }\left( t\right) =\mathbf{z}+\varepsilon ~\mathbf{v}\left(
t\right)   \label{zitazetav}
\end{equation}%
with $\varepsilon $ infinitesimal, and by thereby obtaining from (\ref{DS})
the \textit{linearized} system 
\begin{subequations}
\label{LinSyst}
\begin{equation}
\mathbf{\dot{v}}\left( t\right) =\mathbf{i}~\mathbf{M}\left( \mathbf{z}%
\right) \mathbf{~v}\left( t\right) ~,
\end{equation}%
where of course the ($t$-independent) $N\times N$ matrix $\mathbf{M}\left( 
\mathbf{z}\right) $ is defined componentwise as follows:%
\begin{equation}
M_{nm}\left( \mathbf{z}\right) =\left. \frac{\partial ~f_{n}\left( \mathbf{%
\zeta }\right) }{\partial ~\zeta _{m}}\right\vert _{\mathbf{\zeta =z}}~.
\end{equation}%
Hence the solution of this linearized system (\ref{LinSyst}) is a linear
combination, with constant coefficients, of exponential functions $\exp
\left( \mathbf{i}\lambda _{m}t\right) ,$ where $\lambda _{m}$ are the $N$
eigenvalues of the $N\times N$ matrix $\mathbf{M}\left( \mathbf{z}\right) $.
But if the $t$-evolution of the $N$ solutions $\zeta _{n}\left( t\right) $
of the dynamical system (\ref{DS}) are characterized by the periodicity
property (\ref{T}), the $N$ quantities $v_{n}\left( t\right) $---see (\ref%
{zitazetav})---must also possess this \textit{same} property. Hence the $N$
eigenvalues $\lambda _{m}$ must feature the \textit{Diophantine} property to
be \textit{integers}; and these \textit{integers} can be actually identified
by comparing the behavior of the \textit{solvable} system with that of its 
\textit{linearized }version in the \textit{infinitesimal} and \textit{%
immediate} vicinity of the relevant equilibria.

\textbf{Remark 1.1.} An analogous process can be applied to a \textit{%
second-order} dynamical system $\ddot{\mathbf{\zeta }}=\mathbf{f}(\mathbf{%
\zeta })$ satisfying conditions (i) and (ii), instead of the \textit{%
first-order} system~(\ref{DS}). In this case, the linearization obtained in
the \textit{second step} leads to a \textit{second-order} linear system $%
\ddot{\mathbf{v}}(t)=-\mathbf{M}(\mathbf{z})\mathbf{v}(t)$, where the matrix 
$\mathbf{M}(\mathbf{z})$ is defined componentwise by $M_{nm}\left( \mathbf{z}%
\right) =-\left. \frac{\partial ~f_{n}\left( \mathbf{\zeta }\right) }{%
\partial ~\zeta _{m}}\right\vert _{\mathbf{\zeta =z}}~.$ $\blacksquare $

In this paper we again follow this approach, but with a new twist---based on
a new development allowing to manufacture larger classes of \textit{solvable}
dynamical systems \cite{C2015, BC2015}---yielding results which seem new and
indeed somewhat surprising. Indeed, the $N\times N$ matrix $\mathbf{M}%
^{(1)}\left( \mathbf{z}\right) $, which we identify as having $N$ \textit{%
integer} eigenvalues, is defined in terms of the $N$ \textit{zeros} $z_{n}$
of (monic) polynomials of degree $N$ in $z$ the $N$ \textit{coefficients} of
which are the $N$ \textit{zeros} of Hermite polynomials of degree $N$.
Moreover, by starting from the system of Newtonian (second-order) ODEs the
solvability of which has been demonstrated in \cite{BC2015} rather than from
a first-order system of ODEs such as (\ref{DS}), see \textbf{Remark 1.1}, we
provide an analogous derivation of another $N\times N$ matrix $\mathbf{M}%
^{\left( 2\right) }\left( \mathbf{z}\right) $---again constructed with the
same zeros $z_{n}$---featuring $N$ \textit{squared-integers} as eigenvalues.
These findings are reported in the following Section 2 and proven in Section
3. Section 4 (\textquotedblleft Outlook\textquotedblright ) outlines tersely
some possible further developments.

\section{Results}

In this section we report our findings; their proof is provided in the
following Section 3.

\textbf{Proposition 2.1}. Let $c_{m}$ with $m=1,...,N$ be the $N$ zeros
(arbitrarily ordered) of the Hermite polynomial $H_{N}\left( c\right) $ of
order $N$ and argument $c$ (see for instance \cite{HTF2}): 
\end{subequations}
\begin{subequations}
\begin{equation}
H_{N}\left( c\right) =N!~\sum_{k=0}^{\left[ \left[ N/2\right] \right] }\left[
\frac{\left( -1\right) ^{k}~\left( 2c\right) ^{N-2~k}}{k!~\left(
N-2~k\right) !}\right] ~,
\end{equation}%
\begin{equation}
H_{N}\left( c_{m}\right) =0~,~~~m=1,...,N~,
\end{equation}%
where of course the notation $\left[ \left[ N/2\right] \right] $ denotes the
integer part of $N/2,$ i. e. $N/2$ if $N$ is \textit{even}, $\left(
N-1\right) /2$ if $N$ is \textit{odd}. Let the monic polynomial $p_{N}\left(
z\right) $ of degree $N$ in the variable $z$ be defined as follows: 
\end{subequations}
\begin{equation}
p_{N}\left( z\right) =z^{N}+\sum_{m=1}^{N}\left( c_{m}~z^{N-m}\right)
=\prod\limits_{n=1}^{N}\left( z-z_{n}\right) ~.  \label{pN}
\end{equation}%
The first equality identifies the $N$ \textit{coefficients} $c_{m}$ of this
polynomial $p_{N}\left( z\right) $ as the $N$ \textit{zeros} $c_{m}$ of the
Hermite polynomial of order $N$; note that there are $N!$ such \textit{%
different} polynomials $p_{N}\left( z\right) \equiv p_{N}^{\left( \mu
\right) }\left( z\right) $, $\mu =1,...,N!$, depending on the ordering
assignment of the $N$ zeros $c_{m}$ of the Hermite polynomial of order $N$.
The second equality identifies (\textit{uniquely up to permutations}) the $N$
zeros $z_{n}^{\left( \mu \right) }$ of the polynomial $p_{N}^{\left( \mu
\right) }\left( z\right) $.

Let the $N\times N$ matrices $\mathbf{M}^{(1)(\mu )}\equiv \mathbf{M}%
^{(1)}\left( \mathbf{z}^{\left( \mu \right) }\right) $ be defined
componentwise as follows in terms of the $N$ zeros $\left( z_{1}^{\left( \mu
\right) },\ldots ,z_{N}^{\left( \mu \right) }\right) \equiv \mathbf{z}%
^{\left( \mu \right) }$ of the polynomial $p_{N}^{\left( \mu \right) }\left(
z\right) $: 
\begin{subequations}
\label{M}
\begin{eqnarray}
&&M_{nm}^{(1)\left( \mu \right) }\equiv M_{nm}^{\left( 1\right) }\left( 
\mathbf{z}^{\left( \mu \right) }\right) =-\Bigg\{\prod\limits_{\ell =1,\ell
\neq n}^{N}\frac{1}{[z_{n}^{\left( \mu \right) }-z_{\ell }^{\left( \mu
\right) }]}\Bigg\}  \notag \\
&&\cdot \sum_{j=1}^{N}\Bigg\{\left[ z_{n}^{\left( \mu \right) }\right]
^{N-j}\cdot \Bigg[w_{j,m}^{\left( \mu \right) }+\sum_{s=1,s\neq j}^{N}\frac{%
w_{j,m}^{\left( \mu \right) }-w_{s,m}^{\left( \mu \right) }}{[c_{j}^{\left(
\mu \right) }-c_{s}^{\left( \mu \right) }]^{2}}\Bigg]\Bigg\}~,  \label{Mnm}
\end{eqnarray}%
where 
\begin{equation}
c_{m}^{\left( \mu \right) }\equiv c_{m}\left( \mathbf{z}^{\left( \mu \right)
}\right) ={\left( -1\right) ^{m}}~\sigma _{m}\left( \mathbf{z}^{\left( \mu
\right) }\right)
\end{equation}%
and 
\begin{equation}
w_{j,m}^{\left( \mu \right) }\equiv w_{j,m}\left( \mathbf{z}^{\left( \mu
\right) }\right) ={\left( -1\right) ^{j}}~\left[ \delta _{j1}+\sigma
_{m,j}\left( \mathbf{z}^{\left( \mu \right) }\right) \right]  \label{wjm}
\end{equation}%
(see \textbf{Notation 1.1}).

Then the $N$ eigenvalues $\lambda _{m}^{(1)}$ of the $N\times N$ matrices $%
\mathbf{M}^{(1)\left( \mu \right) }$are given by the following neat (\textit{%
Diophantine}) formula: 
\end{subequations}
\begin{equation}
\lambda _{m}^{(1)}=m~,~~~m=1,2,...,N~.~\blacksquare   \label{Eigen}
\end{equation}

\textbf{Remark 2.1.} Note that in \textbf{Proposition 2.1} \textit{all} the 
\textit{different} $N!$ matrices $\mathbf{M}^{(1)\left( \mu \right) }\equiv 
\mathbf{M}^{(1)}\left( \mathbf{z}^{\left( \mu \right) }\right) $, where $\mu
=1,2,\ldots ,N!$, feature the \textit{same} set of $N$ eigenvalues $%
\lambda^{(1)} _{m}$. Also note that, although definition (\ref{Mnm}) of the
matrix $\mathbf{M}^{(1)\left( \mu \right)}$ depends on the ordering
assignment of the $N$ zeros $z_{n}^{\left( \mu \right) }$ of the polynomial $%
p^{(\mu )}(z),$ such different assignments produce the \textit{same} matrix $%
\mathbf{M}^{(1)\left( \mu \right)}$ up to a \textit{reshuffling of its lines
and columns} and have therefore no relevance for the eigenvalues of $\mathbf{%
M}^{(1)\left( \mu \right)}$: indeed, a switch of the two zeros $z_{n}^{(\mu
)}$ and $z_{m}^{(\mu )}$ of the polynomial $p^{(\mu )}(z)$ yields a new
matrix $\hat{\mathbf{M}}^{(1)(\mu )}$, which corresponds to the original
matrix $\mathbf{M}^{(1)\left( \mu \right) }$ up to the exchange of the $n$%
-th and the $m$-th rows and columns of $\mathbf{M}^{(1)\left( \mu \right) }$
(see also below \textbf{Remark 3.2}). $\blacksquare $

\textbf{Proposition 2.2.} Assume the notation of \textbf{Proposition 2.1}.
If the $N \times N$ matrices $\mathbf{M}^{(2)(\mu)}=\mathbf{M}^{(2)}(\mathbf{%
z}^{\mu})$ are defined by 
\begin{eqnarray}
&&M_{nm}^{\left( 2\right) \left( \mu \right) }\equiv M_{nm}^{\left( 2\right)
}\left( \mathbf{z}^{\left( \mu \right) }\right) =-\Bigg\{\prod\limits_{\ell
=1,\ell \neq n}^{N}\frac{1}{[z_{n}^{\left( \mu \right) }-z_{\ell }^{\left(
\mu \right) }]}\Bigg\}  \notag \\
&&\cdot \sum_{j=1}^{N}\Bigg\{\left[ z_{n}^{\left( \mu \right) }\right]
^{N-j}\cdot \Bigg[w_{j,m}^{\left( \mu \right) }+6~\sum_{s=1,s\neq j}^{N}%
\frac{w_{j,m}^{\left( \mu \right) }-w_{s,m}^{\left( \mu \right) }}{%
[c_{j}^{\left( \mu \right) }-c_{s}^{\left( \mu \right) }]^{4}}\Bigg]\Bigg\},
\label{MnmOld1}
\end{eqnarray}
where $\mu=1,2,\ldots, N!$, then their eigenvalues are given by 
\begin{equation}
\lambda^{(2)}_m=m^2, \;\; m=1,2,\ldots, N.
\end{equation}

\textbf{Remark 2.2.} Note that in \textbf{Proposition 2.2} \textit{all} the 
\textit{different} $N!$ matrices $\mathbf{M}^{(2)\left( \mu \right) }\equiv 
\mathbf{M}^{(2)}\left( \mathbf{z}^{\left( \mu \right) }\right) $, where $\mu
=1,2,\ldots ,N!$, feature the \textit{same} set of $N$ eigenvalues $\lambda
_{m}^{(2)}$. Also note that, although definition (\ref{MnmOld1}) of the
matrix $\mathbf{M}^{\left( 2\right) \left( \mu \right) }$ depends on the
ordering assignment of the $N$ zeros $z_{n}^{\left( \mu \right) }$ of the
polynomial $p^{(\mu )}(z),$ such different assignments produce the \textit{%
same} matrix $\mathbf{M}^{(2)(\mu )}$ up to a \textit{reshuffling of its
lines and columns} and have therefore no relevance for the eigenvalues of $%
\mathbf{M}^{(2)(\mu )}$ (cf. \textbf{Remarks 2.1} and \textbf{3.2}). $%
\blacksquare $

\textbf{Examples 2.1} and \textbf{2.2} illustrate \textbf{Proposition 2.1}
for the cases where $N=2$ and $N=3$.

\textbf{Example 2.1}. Let us construct the $2!=2$ matrices $\mathbf{M}%
^{(1)\left( \mu \right) }$ (see \textbf{Proposition 2.1}) for the case $N=2,$
where the index $\mu \in \{1,2\}$ identifies the $2$ ordering assignments of
the $2$ zeros $\left( c_{1}^{\left( \mu \right) },~c_{2}^{\left( \mu \right)
}\right) =\left( -\frac{\left( -1\right) ^{\mu }}{\sqrt{2}},\frac{\left(
-1\right) ^{\mu }}{\sqrt{2}}\right) $ of the Hermite polynomial of degree $2$%
, $H_{2}(c)=4c^{2}-2$. In this special case the $2$ numbers $c_{1}^{\left(
\mu \right) },~c_{2}^{\left( \mu \right) }$ enter in definition (\ref{Mnm})
of the matrix $\mathbf{M}^{(1)\left( \mu \right) }$ only via the expression $%
[c_{1}^{\left( \mu \right) }-c_{2}^{\left( \mu \right) }]^{-2}=1/2$, which
does not depend on the index $\mu $ distinguishing the two different
orderings of the two zeros $c_{1}^{\left( \mu \right) },~c_{2}^{\left( \mu
\right) }$ of the Hermite polynomial $H_{2}(c)$. Therefore, both choices of
the pair $\Big(c_{1}^{\left( \mu \right) },c_{2}^{\left( \mu \right) }\Big)$
yield the same formula for the matrix $\mathbf{M}^{(1)\left( \mu \right) }$
in terms of the zeros $\mathbf{z}^{\left( \mu \right) }=\Big(z_{1}^{\left(
\mu \right) },z_{2}^{\left( \mu \right) }\Big)$ of the polynomial $%
z^{2}+c_{1}^{\left( \mu \right) }z+c_{2}^{\left( \mu \right) }$: 
\begin{subequations}
\label{MN}
\begin{equation}
\mathbf{M}^{(1)(\mu )}=\mathbf{M}^{(1)}(z_{1}^{(\mu )},z_{2}^{(\mu
)})=\left( 
\begin{array}{cc}
{\ \frac{3}{2}-\frac{1-z_{1}^{\left( \mu \right) }~z_{2}^{\left( \mu \right)
}}{2[z_{1}^{\left( \mu \right) }-z_{2}^{\left( \mu \right) }]}} & -\frac{%
1-[z_{1}^{\left( \mu \right) }]^{2}}{2[z_{1}^{\left( \mu \right)
}-z_{2}^{\left( \mu \right) }]} \\ 
\frac{1-[z_{2}^{\left( \mu \right) }]^{2}}{2[z_{1}^{\left( \mu \right)
}-z_{2}^{\left( \mu \right) }]} & \frac{3}{2}+{\ \frac{1-z_{1}^{\left( \mu
\right) }~z_{2}^{\left( \mu \right) }}{2[z_{1}^{\left( \mu \right)
}-z_{2}^{\left( \mu \right) }]}}%
\end{array}%
\right) ~.  \label{MN2new}
\end{equation}%
Here $z_{1}^{\left( \mu \right) }$ and $z_{2}^{\left( \mu \right) }$ are of
course the $2$ zeros of the polynomial 
\begin{equation}
p^{(\mu )}(z)=z^{2}+\frac{\left( -1\right) ^{\mu }~\left( 1-z\right) }{\sqrt{%
2}}~,~~~\mu =1,2!=1,2~,  \label{p(z)N2}
\end{equation}%
hence%
\begin{equation}
z_{\pm }^{\left( \mu \right) }=\frac{\left( -1\right) ^{\mu }\pm \sqrt{%
1-\left( -1\right) ^{\mu }~4~\sqrt{2}}}{2\sqrt{2}}~,~~\ \mu =1,2
\label{zerosN2}
\end{equation}%
(note that for $\mu =1$ these two zeros are \textit{real}, for $\mu =2$ they
are complex conjugate).

It can be easily verified that each of the matrices $\mathbf{M}^{(1)(\mu )}$%
, $\mu =1,2$, given by formula~(\ref{MN2new}) has the two eigenvalues 1 and
2 consistently with \textbf{Proposition 2.1}. And it is easy to see from
formula (\ref{MN2new}) that the matrix $\mathbf{M}^{(1)}(z_{2}^{(\mu
)},z_{1}^{(\mu )})$ can be obtained from the matrix $\mathbf{M}%
^{(1)}(z_{1}^{(\mu )},z_{2}^{(\mu )})$ by switching its rows and columns,
consistently with \textbf{Remark 2.1.} $\blacksquare $

\textbf{Example 2.2}. For $N=3$ the $3$ zeros of the Hermite polynomial $%
H_{3}\left( c\right) =8c^{3}-12c=4c\left( 2c^{2}-3\right) $ are $0,$ $\pm 
\sqrt{3/2}.$ So we make the following $6=3!$ different assignments: 
\end{subequations}
\begin{subequations}
\label{c1c2c3N3}
\begin{equation}
c_{1}^{\left( 1\right) }=0~,~~~c_{2}^{\left( 1\right) }=\sqrt{3/2}%
~,~~~c_{3}^{\left( 1\right) }=-\sqrt{3/2}~,
\end{equation}%
\begin{equation}
c_{1}^{\left( 2\right) }=0~,~~~c_{2}^{\left( 2\right) }=-\sqrt{3/2}%
~,~~~c_{3}^{\left( 2\right) }=\sqrt{3/2}~,
\end{equation}%
\begin{equation}
c_{1}^{\left( 3\right) }=\sqrt{3/2}~,~~~c_{2}^{\left( 3\right)
}=0~,~~~c_{3}^{\left( 3\right) }=-\sqrt{3/2}~,
\end{equation}%
\begin{equation}
c_{1}^{\left( 4\right) }=\sqrt{3/2}~,~~~c_{2}^{\left( 4\right) }=-\sqrt{3/2}%
~,~~~c_{3}^{\left( 4\right) }=0~,
\end{equation}%
\begin{equation}
c_{1}^{\left( 5\right) }=-\sqrt{3/2}~,~~~c_{2}^{\left( 5\right) }=\sqrt{3/2}%
~,~~~c_{3}^{\left( 5\right) }=0~,
\end{equation}%
\begin{equation}
c_{1}^{\left( 6\right) }=-\sqrt{3/2}~,~~~c_{2}^{\left( 6\right)
}=0~,~~~c_{3}^{\left( 6\right) }=\sqrt{3/2}~;
\end{equation}%
and correspondingly we define the following $6$ polynomials $p_{3}^{\left(
\mu \right) }\left( z\right) $, of third degree in $z$, and their sets of $3$
zeros $z_{1}^{\left( \mu \right) },$ $z_{2}^{\left( \mu \right) },$ $%
z_{3}^{\left( \mu \right) }$: 
\end{subequations}
\begin{equation}
p_{3}^{\left( \mu \right) }\left( z\right) =z^{3}+\sum_{m=1}^{3}\left[
c_{m}^{\left( \mu \right) }~z^{3-m}\right] =\prod\limits_{n=1}^{3}\left[
z-z_{n}^{\left( \mu \right) }\right] ~,~~~\mu =1,...,6~.  \label{p3mu}
\end{equation}%
The $3$ zeros $z_{1}^{\left( \mu \right) },$ $z_{2}^{\left( \mu \right) },$ $%
z_{3}^{\left( \mu \right) }$ can be easily obtained from this formula (if
need be, via Mathematica), but rather than writing here their exact values
(involving square and cubic roots) we simply provide their (of course
approximate) numerical values in decimal form: 
\begin{subequations}
\label{z1z2z3N3}
\begin{equation}
z_{1}^{\left( 1\right) }=0.7090~,~~~z_{2}^{\left( 1\right) }=-0.3545-1.2656~%
\mathbf{i}~,~~~z_{3}^{\left( 1\right) }=-0.3545+1.2656~\mathbf{i}~,
\end{equation}%
\begin{equation}
z_{1}^{\left( 2\right) }=0.7202-0.5758~\mathbf{i}~,~~~z_{2}^{\left( 2\right)
}=0.7202+0.5758~\mathbf{i}~,~~~z_{3}^{\left( 2\right) }=-1.4405~,
\end{equation}%
\begin{equation}
z_{1}^{\left( 3\right) }=-1.0031+0.7492~\mathbf{i}~,~~~z_{2}^{\left(
3\right) }=-1.0031-0.7492~\mathbf{i}~,~~~z_{3}^{\left( 3\right) }=0.7814~,
\end{equation}%
\begin{equation}
z_{1}^{\left( 4\right) }=0~,~~~z_{2}^{\left( 4\right)
}=-1.8772~,~~~z_{3}^{\left( 4\right) }=0.6524~,
\end{equation}%
\begin{equation}
z_{1}^{\left( 5\right) }=0~,~~~z_{2}^{\left( 5\right)
}=-1.8772~,~~~z_{3}^{\left( 5\right) }=0.6524~,
\end{equation}%
\begin{equation}
z_{1}^{\left( 6\right) }=-0.7814~,~~~z_{2}^{\left( 6\right) }=1.0031-0.7492~%
\mathbf{i}~,~~~z_{3}^{\left( 6\right) }=1.0031+0.7492~\mathbf{i}~.
\end{equation}%
Of course these numbers are defined up to permutations, but, as explained
below, this has no relevance for the \textit{eigenvalues }of the matrices $%
\mathbf{M}^{\left( 1\right) }$ and $\mathbf{M}^{\left( 2\right) }$, see 
\textbf{Remarks 2.1 }and \textbf{3.2}. Note moreover that, as it happens,
the zeros of the polynomials $p_{3}^{(4)}(z)$ and $p_{3}^{(5)}(z)$ are the
same.

From formulas (\ref{wjm}) for $N=3$ we moreover obtain that 
\end{subequations}
\begin{equation}
w_{1,m}^{(\mu )}=-1,\;w_{2,m}^{(\mu )}=z_{m+1}^{(\mu )}+z_{m+2}^{(\mu
)},\;w_{3,m}^{(\mu )}=-z_{m+1}^{(\mu )}~z_{m+2}^{(\mu )}~,~~~m=1,2,3~~~\func{%
mod}\left( 3\right) ~.  \label{wjmN3}
\end{equation}

By inserting these values of the quantities $w_{j,m}^{(\mu )}$ into formula (%
\ref{Mnm}) for $N=3$ we obtain the components of the corresponding $6$
matrices $\mathbf{M}^{(1)\left( \mu \right) }$: 
\begin{eqnarray}
&&M_{nm}^{(1)(\mu )}=-\frac{1}{\left[ z_{n}^{(\mu )}-z_{n+1}^{(\mu )}\right]
~\left[ z_{n}^{(\mu )}-z_{n+2}^{(\mu )}\right] }  \notag \\
&&\cdot \Bigg\{-\left[ z_{n}^{(\mu )}\right] ^{2}+z_{n}^{(\mu )}~\left[
z_{m+1}^{(\mu )}+z_{m+2}^{(\mu )}\right] -z_{m+1}^{(\mu )}~z_{m+2}^{(\mu )} 
\notag \\
&& + \left[ z_{n}^{(\mu )}-1\right] ~\left( -\frac{z_{n}^{(\mu )}~\left[
1+z_{m+1}^{(\mu )}+z_{m+2}^{(\mu )}\right] }{\left[ c_{1}^{(\mu
)}-c_{2}^{(\mu )}\right] ^{2}}\right.  \notag \\
&&+\frac{\left[ -1+z_{m+1}^{(\mu )}~z_{m+2}^{(\mu )}\right] ~\left[
z_{n}^{(\mu )}+1\right] }{\left[ c_{1}^{(\mu )}-c_{3}^{(\mu )}\right] ^{2}} 
\notag \\
&&\left. +\frac{\left[ z_{m+1}^{(\mu )}+z_{m+2}^{(\mu )}+z_{m+1}^{(\mu
)}~z_{m+2}^{(\mu )}\right] }{\left[ c_{2}^{(\mu )}-c_{3}^{(\mu )}\right] ^{2}%
}\right) ~\Bigg\}~,~~~m=1,2,3~~~\func{mod}(3)~.  \label{MN3new}
\end{eqnarray}

It can be checked by direct computation that the $6$ matrices given by~(\ref%
{MN3new}) \textit{all} feature the $3$ eigenvalues $1,$ $2,$ $3$. And it is
again plain that a permutation of the zeros of the polynomial $p_{3}^{(\mu
)}(z)$ results in a corresponding permutation of the rows and the columns of
the matrix $\mathbf{M}^{(1)(\mu )}$, see \textbf{Remarks 2.1 }and \textbf{3.2%
}. $\blacksquare $

Let us end this section by displaying the matrices $\mathbf{M}^{\left(
2\right) }\left( \mathbf{z}\right) $, defined in \textbf{Proposition 2.2},
for $N=2$ and $N=3$. Note that these matrices $\mathbf{M}^{\left( 2\right)
}\left( \mathbf{z}\right) $ do \textit{not} coincide with the matrices $%
\left[ \mathbf{M}^{(1)} \left( \mathbf{z}\right) \right] ^{2}$ (see those
reported in \textbf{Examples 2.1} and \textbf{2.2}), although they of course
feature the \textit{same} squared-integer eigenvalues.

\textbf{Example 2.3}. Let us construct the $2!=2$ matrices $\mathbf{M}%
^{(2)\left( \mu \right) }$ for the case $N=2,$ where, as in \textbf{Example
2.1}, the index $\mu \in \{1,2\}$ identifies the $2$ ordering assignments of
the $2$ zeros $\left( c_{1}^{\left( \mu \right) },~c_{2}^{\left( \mu \right)
}\right) =\left( -\frac{\left( -1\right) ^{\mu }}{\sqrt{2}},\frac{\left(
-1\right) ^{\mu }}{\sqrt{2}}\right) $ of the Hermite polynomial of degree $2$%
, $H_{2}(c)=4c^{2}-2$. Similarly to \textbf{Example 2.1}, in this special
case the $2$ numbers $c_{1}^{\left( \mu \right) },~c_{2}^{\left( \mu \right)
}$ enter in definition (\ref{MnmOld1}) of the matrix $\mathbf{M}^{(2)\left(
\mu \right) }$ only via the expression $(c_{1}^{\left( \mu \right)
}-c_{2}^{\left( \mu \right) })^{-4}=1/4$, which does not depend on the index 
$\mu $ distinguishing the two different orderings of the two zeros $%
c_{1}^{\left( \mu \right) },~c_{2}^{\left( \mu \right) }$ of the Hermite
polynomial $H_{2}(c)$. Therefore both choices of the pair $\Big(%
c_{1}^{\left( \mu \right) },c_{2}^{\left( \mu \right) }\Big)$ yield the same
formula for the matrix $\mathbf{M}^{\left( \mu \right) }$ in terms the zeros 
$\mathbf{z}^{\left( \mu \right) }=\Big(z_{1}^{\left( \mu \right)
},z_{2}^{\left( \mu \right) }\Big)$ of the polynomial $z^{2}+c_{1}^{\left(
\mu \right) }z+c_{2}^{\left( \mu \right) }$:%
\begin{equation}
\mathbf{M}^{(2)(\mu )}=\mathbf{M}^{(2)}(z_{1}^{(\mu )},z_{2}^{(\mu
)})=\left( 
\begin{array}{cc}
{\small \frac{5}{2}-\frac{3}{2}\frac{1-z_{1}^{\left( \mu \right)
}~z_{2}^{\left( \mu \right) }}{z_{1}^{\left( \mu \right) }-z_{2}^{\left( \mu
\right) }}} & -\frac{3}{2}\frac{1-\left( z_{1}^{\left( \mu \right) }\right)
^{2}}{z_{1}^{\left( \mu \right) }-z_{2}^{\left( \mu \right) }} \\ 
&  \\ 
\frac{3}{2}\frac{1-\left( z_{2}^{\left( \mu \right) }\right) ^{2}}{%
z_{1}^{\left( \mu \right) }-z_{2}^{\left( \mu \right) }} & \frac{5}{2}+\frac{%
3}{2}{\small \frac{1-z_{1}^{\left( \mu \right) }~z_{2}^{\left( \mu \right) }%
}{z_{1}^{\left( \mu \right) }-z_{2}^{\left( \mu \right) }}}%
\end{array}%
\right) ~.  \label{MN2Old}
\end{equation}%
Here $z_{1}^{\left( \mu \right) }$ and $z_{2}^{\left( \mu \right) }$ are of
course the $2$ zeros of the polynomial (\ref{p(z)N2}) given by (\ref{zerosN2}%
) (note, again, that for $\mu =1$ these two zeros are \textit{real}, for $%
\mu =2$ they are complex conjugate).

It can be easily verified that each of the matrices $\mathbf{M}^{(2)(\mu )}$%
, $\mu =1,2$, given by formula~(\ref{MN2Old}) has the two eigenvalues 1 and
4, consistently with \textbf{Proposition 2.2}. And it is easy to see from
formula (\ref{MN2Old}) that the matrix $\mathbf{M}^{(2)}(z_{2}^{(\mu
)},z_{1}^{(\mu )})$ can be obtained from the matrix $\mathbf{M}%
^{(2)}(z_{1}^{(\mu )},z_{2}^{(\mu )})$ by switching its rows and columns,
consistently with \textbf{Remark 2.2.} $\blacksquare $

\textbf{Example 2.4}. For $N=3$ the $3$ zeros of the Hermite polynomial $%
H_{3}\left( c\right) =8c^{3}-12c=4c\left( 2c^{2}-3\right) $ are $0,$ $\pm 
\sqrt{3/2}$. As in \textbf{Example 2.2}, we make $3!=6$ different
assignments of the coefficients $c_{1}^{(\mu )},c_{2}^{(\mu )},c_{3}^{(\mu )}
$, given by~(\ref{c1c2c3N3}), and define the polynomials $p_{3}^{\left( \mu
\right) }\left( z\right) $ by~(\ref{p3mu}). The $3$ zeros $z_{1}^{\left( \mu
\right) },$ $z_{2}^{\left( \mu \right) },$ $z_{3}^{\left( \mu \right) }$ can
be easily obtained from~(\ref{p3mu}); their approximate numerical values are
given by~(\ref{z1z2z3N3}). Of course these numbers are defined up to
permutations, but, as explained below, this has no relevance for the \textit{%
eigenvalues }of the matrices $\mathbf{M}^{\left( 1\right) }$ and $\mathbf{M}%
^{\left( 2\right) }$, see \textbf{Remark 2.2}.

For $N=3$ the coefficients $w_{j,m}^{(\mu)}$ from (\ref{wjm}) are given by~(%
\ref{wjmN3}). By inserting these values of the quantities $w_{j,m}^{(\mu )}$
into formula (\ref{MnmOld1}) for $N=3$ we obtain the components of the
corresponding $6$ matrices $\mathbf{M}^{(2)\left( \mu \right) }$: 
\begin{eqnarray}
&&M_{nm}^{(2)(\mu )}=-\frac{1}{\left[ z_{n}^{(\mu )}-z_{n+1}^{(\mu )}\right]
~\left[ z_{n}^{(\mu )}-z_{n+2}^{(\mu )}\right] }  \notag \\
&&\cdot \Bigg\{-\left[ z_{n}^{(\mu )}\right] ^{2}+z_{n}^{(\mu )}~\left[
z_{m+1}^{(\mu )}+z_{m+2}^{(\mu )}\right] -z_{m+1}^{(\mu )}~z_{m+2}^{(\mu )} 
\notag \\
&&+6~\left[ z_{n}^{(\mu )}-1\right] ~\left( -\frac{z_{n}^{(\mu )}~\left[
1+z_{m+1}^{(\mu )}+z_{m+2}^{(\mu )}\right] }{\left[ c_{1}^{(\mu
)}-c_{2}^{(\mu )}\right] ^{4}}\right.  \notag \\
&&+\frac{\left[ -1+z_{m+1}^{(\mu )}~z_{m+2}^{(\mu )}\right] ~\left[
z_{n}^{(\mu )}+1\right] }{\left[ c_{1}^{(\mu )}-c_{3}^{(\mu )}\right] ^{4}} 
\notag \\
&&\left. +\frac{\left[ z_{m+1}^{(\mu )}+z_{m+2}^{(\mu )}+z_{m+1}^{(\mu
)}~z_{m+2}^{(\mu )}\right] }{\left[ c_{2}^{(\mu )}-c_{3}^{(\mu )}\right] ^{4}%
}\right) ~\Bigg\}~,~~~m=1,2,3~~~\func{mod}(3)~.  \label{MN3Old}
\end{eqnarray}

It can be checked by direct computation that the $6$ matrices given by~(\ref%
{MN3Old}) \textit{all} feature the $3$ eigenvalues $1,$ $4,$ $9$. Moreover,
it can be verified, again by computation, that a permutation of the zeros of
the polynomial $p_{3}^{(\mu )}(z)$ results in an appropriate permutation of
the rows and the columns of the matrix $\mathbf{M}^{(2)(\mu )}$, see \textbf{%
Remark 2.2 }. $\blacksquare $

\section{Proofs}

In this Section 3 we prove \textbf{Propositions 2.1} and \textbf{2.2}.

\textbf{Proof of Proposition 2.1}. Our starting point is the system of $N$
ODEs 
\begin{subequations}
\label{EqsMotionzitanNew1}
\begin{equation}
\dot{\zeta}_{n}=-\left\{ \left[ \prod\limits_{\ell =1,~\ell \neq
n}^{N}\left( \zeta _{n}-\zeta _{\ell }\right) \right] ^{-1}~\sum_{m=1}^{N}%
\left[ \dot{\gamma}_{m}~\left( \zeta _{n}\right) ^{N-m}\right] \right\}
\label{zitagammadot}
\end{equation}%
(see eq. (9a) of \cite{C2015}), with%
\begin{equation}
\dot{\gamma}_{m}=\mathbf{i~}\left[ \gamma _{m}-\sum_{\ell =1,~\ell \neq
m}^{N}\left( \gamma _{m}-\gamma _{\ell }\right) ^{-1}\right] ~.  \label{CM}
\end{equation}%
Above and hereafter we assume the $2N$ quantities $\zeta _{n}\equiv \zeta
_{n}\left( t\right) $ and $\gamma _{m}\equiv \gamma _{m}\left( t\right) $ to
be functions of the independent variable $t$, and we denote with
superimposed dots differentiations with respect to this variable (the
dependence on which is, for notational simplicity, not always \textit{%
explicitly} displayed, see for instance (\ref{EqsMotionzitanNew1})). We
moreover assume \cite{C2015, BC2015} the quantities $\gamma _{m}\equiv
\gamma _{m}\left( t\right) $ respectively $\zeta _{n}\equiv \zeta _{n}\left(
t\right) $ to be the $N$ \textit{coefficients} respectively the $N$ \textit{%
zeros} of a time-dependent monic polynomial $\psi _{N}\left( \zeta ;t\right) 
$ of degree $N$ in the variable $\zeta $: 
\end{subequations}
\begin{subequations}
\label{psigamma}
\begin{equation}
\psi _{N}\left( \zeta ;t\right) =\zeta ^{N}+\sum_{m=1}^{N}\left[ \gamma
_{m}\left( t\right) ~\zeta ^{N-m}\right] =\prod\limits_{n=1}^{N}\left[ \zeta
-\zeta _{n}\left( t\right) \right] ~,  \label{psiN}
\end{equation}%
implying%
\begin{equation}
\gamma _{m}\left( t\right) ={\left( -1\right) ^{m}}~\left[ \sum_{1\leq
s_{1}<s_{2}<\ldots <s_{m}\leq N}\zeta _{s_{1}}(t)\zeta _{s_{2}}(t)\cdots
\zeta _{s_{m}}(t)\right] ~.  \label{gammazita}
\end{equation}%
It is this connection that justifies the validity of (\ref{zitagammadot}):
see \cite{C2015, BC2015}.

It is on the other hand well known (see for instance \cite{C2001} sect.
2.3.4.1) that dynamical system (\ref{CM}) is \textit{integrable} indeed 
\textit{solvable} and \textit{isochronous} with period $2\pi $, \textit{all}
its solutions featuring the remarkable property 
\end{subequations}
\begin{equation}
\gamma _{m}\left( t+2~\pi \right) =\gamma _{m}\left( t\right) ~.
\label{gammaperiodic}
\end{equation}

Hence, for the dynamical system implied by (\ref{zitagammadot}) with (\ref%
{CM}), reading 
\begin{subequations}
\label{EqsMotionzitanNew2}
\begin{equation}
\dot{\zeta}_{n}=-\mathbf{i~}\left[ \prod\limits_{\ell =1,~\ell \neq
n}^{N}\left( \zeta _{n}-\zeta _{\ell }\right) \right] ^{-1}~\sum_{m=1}^{N}%
\left\{ \left[ \gamma _{m}-\sum_{\ell =1,~\ell \neq m}^{N}\left( \gamma
_{m}-\gamma _{\ell }\right) ^{-1}\right] ~\left( \zeta _{n}\right)
^{N-m}\right\} ~,  \label{Eqzitan}
\end{equation}%
with $\gamma _{m}\equiv \gamma _{m}\left( t\right) $ expressed in terms of
the $N$ zeros $\zeta _{n}\equiv \zeta _{n}\left( t\right) $ via (\ref%
{gammazita}), all the solutions $\zeta _{n}\left( t\right) $ are likewise
periodic with period $2\pi ,$ being the $N$ \textit{zeros} of a polynomial $%
\psi _{N}\left( \zeta ;t\right) $, of degree $N$ in $\zeta $ (see (\ref{psiN}%
)), which is itself periodic with period $2\pi $ since its coefficients are 
\textit{all} periodic with period $2\pi $: 
\begin{equation}
\zeta _{n}\left( t+2~\pi \right) =\zeta _{n}\left( t\right) ~.
\label{zitaperiodic}
\end{equation}

\textbf{Remark 3.1}. It might be observed that the zeros $\zeta _{n}\left(
t\right) $ of a time-dependent polynomial of degree $N$ in $\zeta $ which is
itself periodic with period $T$ might themselves be periodic with a period
which is a (generally small; see \cite{GS2005}) \textit{integer multiple} of 
$T,$ due to the possibility that the zeros, as it were, \textit{exchange
their roles} through their time evolution. But this is certainly not the
case for a time evolution in which each zero evolves periodically remaining
infinitely close to its equilibrium position in an equilibrium configuration
of system (\ref{EqsMotionzitanNew2}); which is the case we consider below. $%
\blacksquare $

It is moreover clear that an equilibrium configuration of this dynamical
system, (\ref{Eqzitan}), is provided by the values $\zeta _{n}\left(
t\right) =z_{n}$ which are the zeros of the polynomial (\ref{psiN})
corresponding to the equilibrium configuration $\gamma _{n}\left( t\right)
=c_{n}$ of the system (\ref{CM}), where the $N$ quantities $c_{n}$ satisfy
the system of algebraic equations 
\end{subequations}
\begin{subequations}
\begin{equation}
c_{m}-\sum_{\ell =1,~\ell \neq m}^{N}\left( c_{m}-c_{\ell }\right)
^{-1}=0~,~\ ~m=1,...,N~.  \label{Eqcm}
\end{equation}%
It is on the other hand well known (see for instance Appendix C of \cite%
{C2001}; this finding is actually much older, see for instance \cite{S1939}
and references therein) that the $N$ zeros $c_{n}$ of the Hermite polynomial
of order $N$ in $c$, 
\begin{equation}
H_{N}\left( c_{n}\right) =0~,~~~n=1,...,N~,
\end{equation}%
satisfy this system of algebraic equations, (\ref{Eqcm}).

Next, let us look at the behavior of dynamical system (\ref%
{EqsMotionzitanNew2}) in the infinitesimal vicinity of the equilibrium
configuration $\zeta _{n}=z_{n}$, where the $N$ coordinates $z_{n}$ are the $%
N$ zeros of the polynomial $p_{N}\left( z\right) ,$ see (\ref{pN}). To this
end we set 
\end{subequations}
\begin{subequations}
\begin{equation}
\zeta _{n}\left( t\right) =z_{n}+\varepsilon ~v_{n}\left( t\right)
~,~~~\gamma _{m}\left( t\right) =c_{m}+\varepsilon ~w_{m}\left( t\right)
+O\left( \varepsilon ^{2}\right) ~,  \label{epsi}
\end{equation}%
with $\varepsilon $ infinitesimal; and we note that, via (\ref{Eqcm}) with (%
\ref{gammazita}), these two formulas, (\ref{epsi}), imply that 
\begin{eqnarray}
&&\gamma _{m}-\sum_{\ell =1,~\ell \neq m}^{N}\left( \gamma _{m}-\gamma
_{\ell }\right) ^{-1}  \notag \\
&=&\varepsilon \left[ w_{m}+\sum_{\ell =1,~\ell \neq m}^{N}\left(
c_{m}-c_{\ell }\right) ^{-2}~\left( w_{m}-w_{\ell }\right) \right]
+O(\varepsilon ^{2})~,
\end{eqnarray}%
where 
\begin{equation}
w_{j}\left( t\right) ={\left( -1\right) ^{j}}~\left\{ \sum_{1\leq
s_{1}<s_{2}<\ldots <s_{j}\leq N}~\left[ \sum_{q=1}^{j}v_{s_{q}}\left(
t\right) \prod\limits_{r=1,~r\neq q}^{j}\left( z_{s_{r}}\right) \right]
\right\} ~.  \label{wmvz}
\end{equation}%
It is then easily seen~-- via these formulas and the insertion of (\ref{epsi}%
) in (\ref{EqsMotionzitanNew2})~-- that the dependent variables $v_{n}\equiv
v_{n}\left( t\right) $ evolve in time according to the following linearized
version of (\ref{EqsMotionzitanNew2}): 
\end{subequations}
\begin{eqnarray}
&&\dot{v}_{n}\left( t\right) =-\mathbf{i~}\left[ \prod\limits_{\ell =1,~\ell
\neq n}^{N}\left( z_{n}-z_{\ell }\right) \right] ^{-1} \\
&&\cdot ~\left\{ \sum_{j=1}^{N}\left( z_{n}\right) ^{N-j}\Big[w_{j}\left(
t\right) +\sum_{\ell =1,~\ell \neq j}^{N}\frac{w_{j}\left( t\right) -w_{\ell
}\left( t\right) }{\left( c_{j}-c_{\ell }\right) ^{2}}\Big]~\right\} ~,
\label{vSystem}
\end{eqnarray}%
where of course the quantities $w_{n}\left( t\right) $ must be replaced by
their expressions (\ref{wmvz}).

Clearly this system can be more compactly rewritten as follows:%
\begin{equation}
\mathbf{\dot{v}}\left( t\right) =\mathbf{i}~\mathbf{M}^{(1)} \mathbf{v}%
\left( t\right) ~,  \label{vSystemShort}
\end{equation}%
with the time-independent $N\times N$ matrix $\mathbf{M}^{(1)}$ defined
componentwise by (\ref{M}).

\textbf{Remark 3.2.} It is easy to see that a switch between the two zeros $%
z_{n}$ and $z_{m}$ of the Hermite polynomial $H_{N}(c)$ in the derivation of
system~(\ref{vSystem}), see~(\ref{epsi}), will result in $v_{n}(t)$ and $%
v_{m}(t)$ exchanging their roles. This, in turn, will result in a switch of
the $n$-th and the $m$-th rows as well as the $n$-th and the $m$-th columns
of the matrix $\mathbf{M}^{(1)}$ in system~(\ref{vSystemShort}), see also~(%
\ref{M}) and \textbf{Remark~2.1}.~$\blacksquare $

System~(\ref{vSystemShort}) implies that the $N$-vector $\mathbf{v}\left(
t\right) $ evolves as follows: 
\begin{subequations}
\begin{equation}
\mathbf{v}\left( t\right) =\sum_{m=1}^{N}\left[ a_{m}~\exp \left( \mathbf{i}%
~\lambda _{m}~t\right) ~\mathbf{u}^{\left( m\right) }\right] ~,
\label{Nv(t)}
\end{equation}%
where $\mathbf{u}^{\left( m\right) }$ respectively $\lambda _{m}$ are the
eigenvectors respectively the eigenvalues of the $N\times N$ matrix $\mathbf{%
M}^{(1)}$ (see (\ref{M})), and the $N$ parameters $a_{m}$ are of course
characterized by the initial values of the $N$-vectors $\mathbf{v}$ so that 
\begin{equation}
\mathbf{v}\left( 0\right) =\sum_{m=1}^{N}\left[ a_{m}~\mathbf{u}^{\left(
m\right) }\right] ~.
\end{equation}%
But we know that the time evolution of the $N$ coordinates $\zeta _{n}\left(
t\right) $ is periodic with period $2\pi ,$ see (\ref{zitaperiodic}), hence
(see the first of the two formulas (\ref{epsi})) the $N$-vector $\mathbf{v}%
\left( t\right) $, of components $v_{n}\left( t\right) ,$ must also be
periodic with period $2\pi .$ Hence---see (\ref{Nv(t)})---the $N$
eigenvalues $\lambda _{m}$ of the $N\times N$ matrix $\mathbf{M}^{(1)}$ must
have the integer values $m^{{}}$ identified in \textbf{Proposition 2.1}, see
(\ref{Eigen}). Q. E. D.

\textbf{Proof of Proposition 2.2}. Here we derive an analogous result to
that reported in \textbf{Proposition 1.1}---identifying thereby a $N\times N$
matrix $\mathbf{M}^{\left( 2\right) }\left( \mathbf{z}\right) $ featuring as
its $N$ eigenvalues the \textit{squared integers} $m^{2}$. The derivation of
this finding---which was actually the first one we obtained, as follow-up to 
\cite{BC2015}---is reported below rather tersely, since it is quite
analogous to the proof of \textbf{Proposition 1.1}, see above. Also the
notation is generally identical to that used above, although the quantities
used below should not be identified with those used above; only in some case
we have appended the upper symbol $^{\left( 2\right) },$ to emphasize the
difference of the results reported below from those of \textbf{Proposition
2.1} and its proof.

Our starting point is the novel dynamical system characterized by the $N$
Newtonian equations of motion of goldfish type (see \cite{BC2015}) 
\end{subequations}
\begin{subequations}
\label{EqsMotionzitanOld}
\begin{eqnarray}
&&\ddot{\zeta}_{n}=\sum_{\ell =1,~\ell \neq n}^{N}\left( \frac{2~\dot{\zeta}%
_{n}~\dot{\zeta}_{\ell }}{\zeta _{n}-\zeta _{\ell }}\right)  \notag \\
&&-\left\{ \left[ \prod\limits_{\ell =1,~\ell \neq n}^{N}\left( \zeta
_{n}-\zeta _{\ell }\right) \right] ^{-1}~\sum_{m=1}^{N}\left[ \ddot{\gamma}%
_{m}~\left( \zeta _{n}\right) ^{N-m}\right] \right\} ~,
\label{zitagammadotdot}
\end{eqnarray}%
with%
\begin{equation}
\ddot{\gamma}_{m}=-\gamma _{m}+2~\sum_{\ell =1,~\ell \neq m}^{N}\left(
\gamma _{m}-\gamma _{\ell }\right) ^{-3}~.  \label{CCMMOld}
\end{equation}
We again assume, as above, that the quantities $\zeta _{n}\equiv \zeta
_{n}\left( t\right) $ respectively $\gamma _{m}\equiv \gamma _{m}\left(
t\right) $ are the $N$ \textit{zeros} respectively the $N$ \textit{%
coefficients} of a time-dependent monic polynomial $\psi _{N}\left( \zeta
;t\right) $ of degree $N$ in the variable $\zeta ,$ see (\ref{psigamma}).

It is on the other hand well known that dynamical system (\ref{CCMMOld}) is 
\textit{integrable} indeed \textit{solvable} and \textit{isochronous} with
period\textit{\ }$2\pi $, \textit{all} its solutions featuring the remarkable%
\textit{\ }property 
\end{subequations}
\begin{equation}
\gamma _{m}\left( t+2~\pi \right) =\gamma _{m}\left( t\right)
\end{equation}%
(see for instance \cite{BC2015, C2001, C2008}). Hence, for the dynamical
system implied by (\ref{zitagammadotdot}) with (\ref{CCMMOld}), reading 
\begin{eqnarray}
&&\ddot{\zeta}_{n}=\sum_{\ell =1,~\ell \neq n}^{N}\left( \frac{2~\dot{\zeta}%
_{n}~\dot{\zeta}_{\ell }}{\zeta _{n}-\zeta _{\ell }}\right) -\left[
\prod\limits_{\ell =1,~\ell \neq n}^{N}\left( \zeta _{n}-\zeta _{\ell
}\right) \right] ^{-1}  \notag \\
&&\cdot \sum_{m=1}^{N}\left\{ \left[ -\gamma _{m}+2~\sum_{\ell =1,~\ell \neq
m}^{N}\left( \gamma _{m}-\gamma _{\ell }\right) ^{-3}\right] ~\left( \zeta
_{n}\right) ^{N-m}\right\} ~,  \label{DynamSystzita}
\end{eqnarray}%
with $\gamma _{m}\equiv \gamma _{m}\left( t\right) $ expressed in terms of
the $N$ zeros $\zeta _{n}\equiv \zeta _{n}\left( t\right) $ via (\ref%
{gammazita}), all the solutions $\zeta _{n}\left( t\right) $ are likewise
periodic with period $2\pi ,$ being the $N$ \textit{zeros} of a polynomial $%
\psi _{N}\left( \zeta ;t\right) $, of degree $N$ in $\zeta $ (see (\ref{psiN}%
)), the coefficients of which are \textit{all} periodic with period $2\pi $:%
\begin{equation}
\zeta _{n}\left( t+2~\pi \right) =\zeta _{n}\left( t\right) ~.
\label{zitaperiodicOld}
\end{equation}

Indeed an equilibrium configuration of dynamical system (\ref{DynamSystzita}%
) is provided by the $N$ zeros $z_{n}$ of a polynomial $p_{N}\left( z\right) 
$ the $N$ coefficients $c_{m}$ of which satisfy the set of $N$ algebraic
equations%
\begin{equation}
-c_{m}+2~\sum_{\ell =1,~\ell \neq m}^{N}\left( c_{m}-c_{\ell }\right)
^{-3}=0~;  \label{Equicm}
\end{equation}%
since clearly the right-hand sides of the $N$ ODEs (\ref{DynamSystzita}) all
vanish at equilibrium---i. e. when all the \textquotedblleft
velocities\textquotedblright\ $\dot{\zeta}_{n}$ vanish---provided moreover $%
\gamma _{m}=c_{m},$ $m=1,...,N$ with the $N$ parameters $c_{m}$ satisfying
the $N$ algebraic equations (\ref{Equicm}). But it is well known (see for
instance \cite{ABCOP1979, C2001}) that the $N$ zeros $c_{n}$ of the Hermite
polynomial $H_{N}\left( c\right) $ do satisfy the set of $N$ algebraic
equations (\ref{Equicm}) (in addition to satisfying the different set of $N$
algebraic equations (\ref{Eqcm})).

Next, let us look at the behavior of dynamical system (\ref{DynamSystzita})
in the infinitesimal vicinity of the equilibrium configuration $\zeta
_{n}=z_{n}$, where the $N$ coordinates $z_{n}$ are the $N$ zeros of the
polynomial $p_{N}\left( z\right) ,$ see (\ref{pN}). Then via (\ref{epsi}) we
get%
\begin{eqnarray}
&&-\gamma _{m}+2~\sum_{\ell =1,~\ell \neq m}^{N}\left( \gamma _{m}-\gamma
_{\ell }\right) ^{-3}  \notag \\
&=&-\varepsilon \left[ w_{m}+6~\sum_{\ell =1,~\ell \neq m}^{N}\left(
c_{m}-c_{\ell }\right) ^{-4}~\left( w_{m}-w_{\ell }\right) \right]
+O(\varepsilon^2)~,
\end{eqnarray}%
of course with $w_{j}\left( t\right) $ defined by (\ref{wmvz}).

It is then easily seen---via these formulas---that the insertion of (\ref%
{epsi}) in (\ref{DynamSystzita}) implies that the dependent variables $%
v_{n}\equiv v_{n}\left( t\right) $ evolve in time according to the following
linearized version of (\ref{DynamSystzita}): 
\begin{eqnarray}
&&\ddot{v}_{n}^{\left( 2\right) }\left( t\right) =-\left[ \prod\limits_{\ell
=1,~\ell \neq n}^{N}\left( z_{n}-z_{\ell }\right) \right] ^{-1}  \notag \\
&&\cdot \sum_{j=1}^{N} \left( z_{n}\right) ^{N-j} \left\{ w_{j}\left(
t\right) +6~\sum_{\ell =1,~\ell \neq j}^{N}\frac{w_{j}\left( t\right)
-w_{\ell }\left( t\right)}{\left( c_{j}-c_{\ell }\right) ^{4}} ~\right\}~,
\end{eqnarray}%
where the quantities $w_{n}\left( t\right) $ must be replaced by their
expressions (\ref{wmvz}) (of course with $v_{m}\left( t\right) $ replaced by 
$v_{m}^{\left( 2\right) }\left( t\right) $).

Clearly this system can be more compactly rewritten as follows:%
\begin{equation}
\mathbf{\ddot{v}}^{\left( 2\right) }\left( t\right) =-\mathbf{M}^{\left(
2\right) }\mathbf{~v}^{\left( 2\right) }\left( t\right) ~,
\label{vSystemShortOld}
\end{equation}%
with the time-independent $N\times N$ matrix $\mathbf{M}^{\left( 2\right) }$
defined by~(\ref{MnmOld1}).

System~(\ref{vSystemShortOld}) implies that the $N$-vector $\mathbf{v}\left(
t\right) $ evolves as follows: 
\begin{subequations}
\begin{equation}
\mathbf{v}^{\left( 2\right) }\left( t\right) =\sum_{m=1}^{N}\left\{ \left[
a_{m}^{\left( 2\right) }~\cos \left( \lambda _{m}^{\left( 2\right)
}~t\right) +b_{m}^{\left( 2\right) }~\frac{\sin \left( \lambda _{m}^{\left(
2\right) }~t\right) }{\lambda _{m}^{\left( 2\right) }}\right] ~\mathbf{u}%
^{\left( 2\right) \left( m\right) }\right\} ~,  \label{Nv(t)Old}
\end{equation}%
where $\mathbf{u}^{\left( 2\right) \left( m\right) }$ respectively $\left[
\lambda _{m}^{\left( 2\right) }\right] $ are the eigenvectors respectively
the eigenvalues of the $N\times N$ matrix $\mathbf{M}^{\left( 2\right) }$,
and the $2N$ parameters $a_{m}^{\left( 2\right) }$ and $b_{m}^{\left(
2\right) }$ are of course characterized by the initial values of the two $N$%
-vectors $\mathbf{v}$ and $\mathbf{\dot{v}}$ so that 
\begin{equation}
\mathbf{v}^{\left( 2\right) }\left( 0\right) =\sum_{m=1}^{N}\left[
a_{m}^{\left( 2\right) }~\mathbf{u}^{\left( 2\right) \left( m\right) }\right]
~,~~~\mathbf{\dot{v}}^{\left( 2\right) }\left( 0\right) =\sum_{m=1}^{N}\left[
b_{m}^{\left( 2\right) }~\mathbf{u}^{\left( 2\right) \left( m\right) }\right]
~.
\end{equation}%
But we know that the time evolution of the $N$ coordinates $\zeta _{n}\left(
t\right) $ is periodic with period $2\pi ,$ see (\ref{zitaperiodicOld}),
hence (see the first of the two formulas (\ref{epsi})) the $N$-vector $%
\mathbf{v}\left( t\right) $, of components $v_{n}\left( t\right) ,$ must
also be periodic with period $2\pi .$ Hence---see (\ref{Nv(t)Old})---the $N$
eigenvalues $\lambda _{m}^{\left( 2\right) }$ of the $N\times N$ matrix $%
\mathbf{M}^{\left( 2\right) }\left( \mathbf{z}\right) $ must have the
squared-integer values $m^{2}$, $m=1,2,\ldots, N$. Q.E.D.

\section{Outlook}

The findings reported in this paper suggest some further developments, which
we intend to pursue in future publications. A quite natural one is the
extension of these results to other, more general, classes of polynomials
than Hermite polynomials (see, for instance, the properties of the zeros of
polynomials reported in \cite{ABCOP1979, BC}). Another avenue of extension
that we plan to explore is by iterating the approach used in this paper, to
find properties of the \textit{zeros} of monic polynomials the \textit{%
coefficients} of which are the \textit{zeros} of monic polynomials the 
\textit{coefficients} of which are the \textit{zeros} of, say, Hermite
polynomials; with an obviously ample choice of how this procedure can be
further iterated and also combined with findings involving other named
polynomials than Hermite polynomials.

\end{subequations}

\end{document}